\documentclass[paper=a4, fontsize=11pt]{article}

\usepackage[table]{xcolor}
\usepackage{color,colortbl,tabularx}
\usepackage[toc,page]{appendix}
\usepackage[english]{babel}
\usepackage[protrusion=true,expansion=true]{microtype}
\usepackage{amsmath,amsfonts,amsthm}
\usepackage[pdftex]{graphicx}
\usepackage{pifont}
\usepackage{soul}

\usepackage[hang,small,labelfont=bf,up,textfont=it,up]{caption}
\usepackage{epstopdf}
\usepackage{subfig}
\usepackage{booktabs}
\usepackage[nodayofweek]{datetime}

\usepackage{hyperref}
\usepackage{url}
\usepackage{float}
\usepackage{amsthm}
\usepackage{courier}
\usepackage{listings}  
\usepackage{multirow}
\usepackage{booktabs}
\usepackage{color}
\usepackage{longtable}

\usepackage{lineno}

\usepackage[margin=2.2cm]{geometry}

\graphicspath{{images/}}

\def\black{\color{black}}

\definecolor{LightBlue}{rgb}{0.88,0.9,0.9}
\usepackage{verbatim}
\usepackage{fancyvrb}
\DefineShortVerb{\|}

\usepackage{sectsty}
\allsectionsfont{
	\usefont{OT1}{bch}{b}{n}
	}

\sectionfont{
	\usefont{OT1}{bch}{b}{n}
	\sectionrule{0pt}{0pt}{-5pt}{0.8pt}
	}
    
\newcommand{\beginsupplement}{%
        \setcounter{table}{0}
        \renewcommand{\thetable}{S\arabic{table}}%
        \setcounter{figure}{0}
        \renewcommand{\thefigure}{S\arabic{figure}}%
        \setcounter{section}{0}
        \renewcommand{\thesection}{S\arabic{section}}%
     }

\usepackage{fancyhdr}
\title{ \vspace{-0.5in} 	\usefont{OT1}{bch}{b}{n}
		\huge \strut EHRs Data Harmonization Platform, an easy-to-use shiny app based on \emph{recodeflow} for harmonizing and deriving clinical features
\strut \\
}
\author{ 									
    \usefont{OT1}{bch}{m}{n}
    \Large Arian Aminoleslami$^1$  \\
    \normalsize $^1$University of Toronto, \\
    \normalsize Toronto, Ontario, Canada \\
    \normalsize $^1$ORCID: {0000-0002-9221-1897}  \\
    \and
    \usefont{OT1}{bch}{m}{n}
    \Large Geoffrey M. Anderson$^2$  \\
    \normalsize $^2$University of Toronto, \\
    \normalsize Toronto, Ontario, Canada \\
    \small $^2$ORCID:~{0000-0003-0124-1343} \\
    \and
    \usefont{OT1}{bch}{m}{n}
    \Large Davide Chicco$^3$\thanks{corresponding author: Davide~Chicco, Universit\`a di Milano-Bicocca (Milan, Italy) \& University of Toronto (Toronto, Ontario, Canada). Email address: \href{mailto:davidechicco@davidechicco.it}{davidechicco@davidechicco.it}}  \\
    \normalsize $^3$Universit\`a di Milano-Bicocca \\
    \normalsize Milan, Italy \\
    \normalsize \& University of Toronto, \\
    \normalsize Toronto, Ontario, Canada \\
    \normalsize $^3$ORCID: {0000-0001-9655-7142} \\
}
\begin{document}
\maketitle

%

\abstract{
Electronic health records (EHRs) contain important longitudinal information on individuals who have received medical care. Traditionally, EHRs have been used to support a wide range of administrative activities such as billing and clinical workflow, but, given the depth and breadth of clinical and demographic data they contain, they are increasingly being used to provide real-world data for research.  Although EHR data have enormous research potential, the full realization of that potential requires a data management strategy that extracts from large EHR databases, that are collected from a range of care settings and time periods, well-documented research-relevant data that can be used by different researchers. Having a common well-documented data management strategy for EHR will support reproducible research and sharing documentation on research variables that are derived from EHR variables is important to open science. In this short paper, we describe the \emph{EHRs Data Harmonization Platform}.  The platform is based on an easy to use web app a publicly available at \url{https://poxotn-arian-aminoleslami.shinyapps.io/Arian/} and as a standalone software package at \url{https://github.com/ArianAminoleslami/EHRs-Data-Harmonization-Platform}, that is linked to an existing R library for data harmonization called \emph{recodeflow}. 
The platform can be used to extract, document, and harmonize variables from EHR and it can also be used to document and share research variables that have been derived from those EHR data.  We provide an example of how this platform is being used to create an environment where multiple research teams can access well-documented data from multiple EHR data sources to conduct reproducible research and where they can share research data sets and the research variables they have created to support open science. We also use a publicly available data set to demonstrate some of the key functions of the platform. We believe this platform can support the use of EHR data for high quality, reproducible research. }


\hspace{0.5cm}


\newpage

\section{Introduction}
\label{sec:INTRODUCTION}

\textbf{Electronic health records}.
Electronic health records~(EHRs) are information systems that collect, store and provide data on healthcare to individuals in digital or electronic format \cite{kim2019evolving}. EHR data are typically collected at the time of healthcare encounters or when services are provided to individuals. These EHR data are collected in multiple care settings and across time and can be linked at the individual level to provide an aggregate description of care for an individual across time and service provider. 
Historically, these data have been used to support administrative functions such as billing, clinical information sharing and workflow management, but EHRs are increasingly being used to support health and health services research \cite{cowie2017electronic,chicco2020survival}.

The data in EHRs typically include basic demographic facts such as date of birth and sex at birth, as well a data on diagnoses, treatments, results of clinical investigations, and outcomes such as hospital discharge or death. Information in EHR is collected at the individual level and using unique patient identifiers it is possible to create trajectories of multiple aspects of care and outcomes over time. These real-world data from EHRs could have enormous value to researchers who are interested in understanding health and disease, in identifying disease trajectories and evaluating disease risks and treatments using modern statistical and machine learning techniques \cite{cowie2017electronic}

\textbf{Using electronic health records for research}.
The process of making EHR data available for research can be divided into two main steps. 
The first step is the selection and description of the specific variables that the data custodian or owner makes available from the EHRs for research purposes and the second step is the process through which individual research teams select and modify variables that are made available to them. 
The first step allows the data custodians to select data elements that they feel are sufficiently well characterized and accurate enough to support the scientific research and importantly allows them to explicitly deal with concerns about protecting the privacy of individuals whose person health information is contained in the EHRs~\cite{fortier2017maelstrom}. 
Typically, this aspect limits access to data elements such as date of birth or address that could identify individuals but can be extended to cover opportunities for re-identification through several variables none of which individually is identifying but that can be in  combination.  
This process typically results in the creation of data dictionaries that provide labels and descriptions of the available EHR data elements.  
The data are often provided in different files from different care settings with different dictionaries. 
For example, hospital inpatient data in one file and outpatient prescription drug data in another file.

The second step in using EHR data for scientific purposes involves selecting the variables from those data sources that are relevant to a specific research project, and creating or deriving research study variables from the available data.  
This step may vary from research team to research team.  
For example, a research team focused on breast cancer might focus on using diagnostic codes from hospital inpatient data that identify individuals with different types of breast cancer and then linking to outpatient prescription drug  data to identify the specific breast cancer treatments they receive, while a research team looking at emergency room waiting times might focus on data from emergency rooms and define waiting time as the time between registration and assessment. Both teams are drawing from the same publicly available data sources but are drawing on different data elements and creating or deriving research variables from those data elements. An important aspect of multiple research teams making use of common data sources is open science, the notion that science will progress faster if data and knowledge are shared \cite{burgelman2019open}.

We have developed a software platform that can support open science in the context of research using publicly available EHR data.
The software platform draws on the data dictionaries that are provided to researchers who are accessing EHR data as the input to a process that allows different research teams to use a similar approach to documentation and harmonization of multiple EHR data sources so that the process for data selection and variable derivation can be standardized and shared. \black 

\textbf{EHRs Data Harmonization Platform}.
The EHRs Data Harmonization Platform is an easy to use publicly available Shiny app that draws on an existing R library: \textit{recodeflow}. 
The R library \textit{recodeflow} was developed as an extension of \emph{cchsflow}\cite{yusuf2021cchsflow,WWW-CCHSFLOW} and itself relies on \textit{sjmisc} \cite{ludecke2018sjmisc}. 
The platform creates shareable documentation of EHR data extraction and derivation that can not only support efforts to make research reproducible, but also will allow researchers to share strategies for data extraction and variable derivation. 

\textbf{\textit{recodeflow} and its components}.
Doug Manuel and colleagues~\cite{yusuf2021cchsflow} originally developed \textit{cchsflow} as a standalone R package and they published it in the Comprehensive R Archive Network~(CRAN)~\cite{WWW-CCHSFLOW}.
Starting from that specific case focused on Canadian Community Health Survey~(CCHS) surveys' data, the same team developed a general version of this software library, called \textit{recodeflow}, that is intended for any survey dataset and any EHR dataset.
The authors released this R library in CRAN, too~\cite{WWW-RECODEFLOW}.
\textit{recodeflow} package is designed to enhance the reproducibility and standardization of data recoding processes. It leverages two critical components: the ``variable details sheet'' and the ``variable sheet''. 

The ``variable sheet'' serves as a repository for essential metadata, including variable labels, which are crucial for accurately interpreting data. 
This sheet ensures that descriptive information about each variable is consistently applied throughout the analysis, promoting clarity and consistency. 
The ``variable details sheet'' also plays a pivotal role in recoding the dataset, as it contains valuable information about variable categories.

By harnessing the insights stored in these two spreadsheets, \textit{recodeflow} empowers researchers to systematically and transparently recode their data, thereby enhancing the rigor and reproducibility of their analytical workflows. 

 We provide a description of essential information stored in each of the spreadsheets along with examples showing how the spreadsheets are filled and should be interpreted in Supplementary~\autoref{fig:DESCRIPTION} and ~\autoref{fig:VARIABLE-SHEET}.
 
\textbf{How does the platform utilize \textit{recodeflow} and contributes in EHRs data harmonization?}
Doug Manuel and his team built \textit{chsflow} and \textit{recodeflow} mainly on \textit{sjmisc}~\cite{WWW-SJMISC}, a popular R package aimed at recoding, dichotomizing or grouping variables, and for general data transformation~\cite{ludecke2018sjmisc}.
\texttt{sjmisc} is available both in CRAN and in Conda~\cite{gruning2018bioconda}.
\textit{recodeflow }performs variable harmonization and derivation, but does not provide a user interface: only software developers proficient in R can use it effectively.

Our platform not only helps in creating the above-mentioned spreadsheets in a user-friendly environment, but also gives the opportunity to users to implement the recoding process on their datasets by taking simple steps. It also documents all essential information (such as the functions’ codes to create the derived variables and their names) and therefore, other researchers can reproduce an already existing work by only uploading the required documentation to the app. 

To be more specific, non-recoded data can be imported to the app with various format such as CSV, SAS7BDAT, RDS, and SQLite.  
There are also options to handle large datasets to be imported to the app in smaller chunks. Users can create a details sheet from scratch using the basic transformations available in \textit{recodeflow} (for example, renaming a variable, creating a categories out of a continuous variable, etc.) or creating more complicated derived variables that has more than one components and needs functions to be coded. The platform then uses the information stored in the created spreadsheets to perform the curation on the dataset. The advantage of this standard approach is that once other users want to perform the same curation on a dataset, they don’t need to create everything from scratch. These spreadsheets could be shared with other users, and they can upload them to the platform, modify them if needed, connect their non-curated database and reproduce the same curation on their database. The platform gives the flexibility to the users to save the curated database in various formats.


In line with the principles of open science~\cite{kush2020fair}, a range of R software libraries and programs are available for data harmonization. 
\emph{retroharmonize}, specializes in survey data harmonization by creating a reusable metadata object that supports the generation of data dictionary, amongst other applications\cite{WWW-RETROHARMONIZE, WWW-RETROHARMONIZE-CRAN}. 
\emph{DataHarmonizer} is an online platform that was created for genomics data but it can be used for other data. 
\emph{DataHarmonizer}~\cite{gill2023dataharmonizer} builds on a new emerging open data model, \emph{LinkML} that supports ontologies and transformations to other databases~\cite{WWW-LINKML}. 
\emph{dxpr} integrates and preprocessing electronic health records, focusing on diagnostic and procedure codes~\cite{tseng2021dxpr}. 
Our EHR Data Harmonization Platform aims to make data transformation as easy as possible with a web-based interface to facilitate tasks that are common for all three applications, including data loading, common transformations, facilitating the generation and organization of R code for new, more complex derived transformation, and display of summary data and metadata. 
Other software programs aimed at harmonizing EHRs~data exist~\cite{zhou2022multiview,heavner2023path,kush2020fair}, but none of them has become a standard tool in the medical environments worldwide.

\textbf{This study}.
We organize the rest of this manuscript as follows.
After this Introduction, we first provide a case study of the use of the platform to support multiple research teams accessing a new EHR data resource for COVID-19 pandemic research~(\autoref{sec:METHODS}). 
We then use a publicly available data set to demonstrate specific functions of the platform~(\autoref{sec:RESULTS}).
Finally, we outline a discussion and some conclusions in the section~(\autoref{sec:DISCUSSION-AND-CONCLUSIONS}).

\section{Methods: Platform's application}\label{sec:METHODS}

In this section, we explain the whole process of our harmonization platform by describing the complete workflow of its application as a tool for data curation and harmonization used by multiple independent research teams making use of a new EHR access environment.  As part of its response to the COVID-19 pandemic, the Ontario provincial government, as the custodian of a large number of EHRs, established a new process through which research teams could access the EHR data in a secure high performance computing environment to conduct REB-approved research on the COVID-19 pandemic~\cite{WWW-OHDP}. 
The Ontario Health Data Platform at Queen’s (OHDP-Q) provided these researchers with access to an array of different types of EHR data and data dictionaries for each of these sources~\cite{WWW-OHDP-DATASETS}. 


These EHRs data came from a range of settings including hospital care~(Discharge Abstract Database, DAD), emergency room care~(National Ambulatory Care Reporting System, NACRS), drug prescriptions~(Ontario Drug Benefit, ODB) and physician visits~(Ontario Health Insurance Plan, OHIP) collected over a 15-year period. 
These data came with data dictionaries for each source and meta data that described changes in variable descriptions for each source over time. 
Each research team conducting research in OHDP-Q had to submit a detailed research plan to the provincial government outlining their project and the EHR files that they would use.  
The provincial government approved projects and the research teams agreed to adhere to policies around data access, privacy, and confidentiality.
 
\begin{figure}[!httb]
\begin{center}
\includegraphics[width=0.4\textwidth]{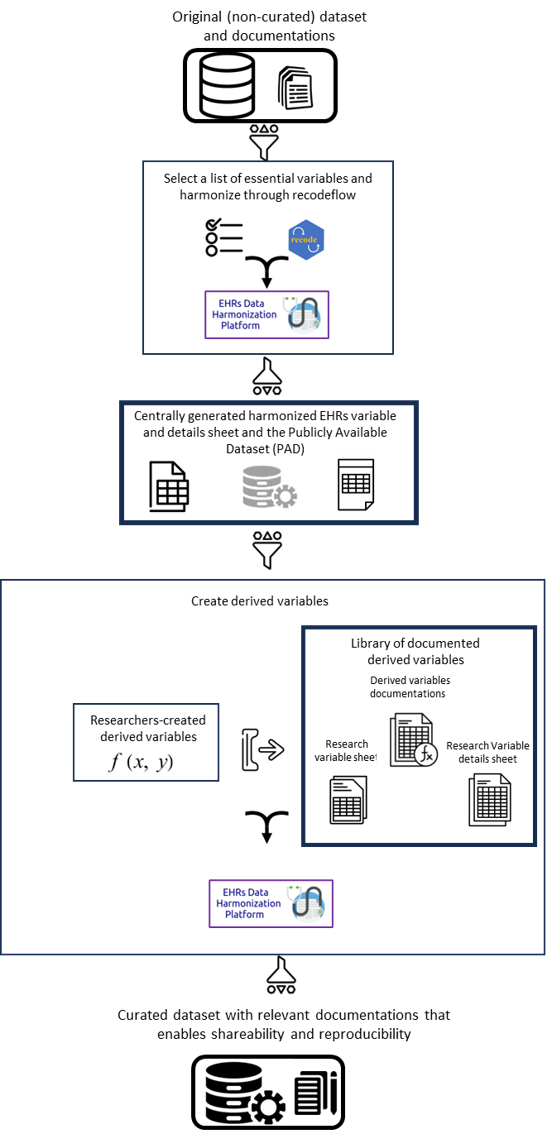}
\caption{Flowchart representing how we utilized EHRs data curation platform to create a standard approach for data curation in OHDP-Q}.
\label{fig:FLOWCHART}
\end{center}
\end{figure}

\paragraph{Use of the EHR Data Harmonization Platform in OHDP-Q for EHR data extraction and harmonization}
We agreed to support the creation of curated data for the research teams using the platform.  
We met with the research support team and the research teams and with them we identified the variables that would be included in the curated data that would be available to the research team.  
Some variables such as six-character postal code and full date of birth (that is, YYYY-MM-DD) were not extracted from the data for privacy reasons and replaced with less specific variables (that is, year of birth, first three characters of postal code). 
The researchers came to agreement on a set of core variables they wanted from each EHR source.  
The platform was then used to select these variables and to recode the postal code and year of birth variables to their new formats.  
There were changes in some of the variable names and in the coding for categorical variables over time.  These changes were incorporated, and data were harmonized over time.  
It was also agreed that the variable names would be changed from those provided in the sources EHR files to names for those variables that were commonly used by researchers in the province.  
The end result of the process was a set of variable and variable details sheets that were used by the research support firm as inputs to the \textit{recodeflow} process that when applied to the source EHR data produced the curated and documented data sets that were made available to the research teams data sets. 
The variable and variable details sheets were provided to the research team as data dictionaries for those curated data sets. 

\paragraph{Use of the EHR Data Harmonization Platform in OHDP-Q for documentation of derived variables}
The research teams in OHDP-Q are working in a common environment that allowed each team access to all the harmonized data sets for which they were approved. In this environment, each team can use the platform to select variables the team wants from each of these harmonized data sources and use those variables to create derived research variables.  A derived variable in its simplest form can be constructed from a single harmonized variable. For example, a researcher team interested in identifying patients with breast cancer treated in hospital could use the variable that describes the diagnose for hospitalized patients from DAD using ICD-10 codes and then create a variable breast\_cancer that is 1 if the ICD-10 code is C50 and 0 if any other diagnosis.  Similarly, the research team might be interested in treatment of patients with a specific drug, say tamoxifen, and they could create a tamoxifen treatment variable that would take a value of 1 if patient was prescribed tamoxifen and 0 if not taking that drug using the ODB prescription data. A more complex derived variable can be created from two or more variables.  For example, the team might want to identify breast cancer patients who were prescribed tamoxifen and could derive that variable from the previous two variables.  The key feature is that derived variables are based on one or more existing variables based on specific rules or logic.  The documentation of derived variables requires the names of the variables that are used in the creation of the derived variables and the function or rule that creates the derived variable.  The platform has a feature that creates and documents derived variables. It is possible to share this documentation.  Using the platform, research teams can create and document the derived variables they are using.  They can then make that documentation available to other research teams.  We have set up a derived variable library where research teams can store the documentation for the variables they create.  If other researchers use as the basis for their documentation the same variable and variable detail sheets that define the shared curated data, they can use the platform to take the derived variable documentation from the library and apply it to their data. 
\autoref{fig:FLOWCHART} illustrates our approach in OHDP-Q

\section{Results: use cases}
\label{sec:RESULTS}

In this section, we’ll show how the platform works by using it on the Paquid dataset of mental health records~\cite{letenneur1994incidence,WWW-PAQUID}, publicly available within the \texttt{lcmm} R package on extended mixed models~\cite{proust2015estimation}.
This Paquid dataset contains 2,250 observations (represented as rows in the dataset table) over 500 subjects and 12 variables (represented as columns in the dataset table).
A patient can have multiple observations.

The variables include repeated cognitive measures (MMSE, IST, and BVRT psychometric tests), physical dependency (HIER) and depression sympatomatology (CESD) collected over 20 years along with dementia information (age at dementia diagnosis, dementia diagnosis information) and time-independent socio-demographic information (CEP, male sex, initial age)~\cite{letenneur1994incidence,WWW-PAQUID}.
This dataset has 726 missing values, that is 2.69\% of the total data instances.

\subsection{Launching the Shiny app}
\label{sec:LAUNCH}
Our Shiny app can be used via web browser by simply visiting the following website: \\
\url{https://poxotn-arian-aminoleslami.shinyapps.io/Arian/}

Alternatively, users can decide to run the Shiny app on their personal computers.
In this case, we report here the instructions for computers using RStudio:
\begin{enumerate}
    \item Install RStudio~\cite{allaire2012rstudio} on your computer;
    \item Download the GitHub repository from \href{https://github.com/ArianAminoleslami/EHRs-Data-Harmonization-Platform}{the URL we indicated}, by clicking first on ``Code'' and then on ``Download ZIP'', or by executing the following command in the shell terminal in Linux systems: \texttt{git clone https://github.com/ArianAminoleslami/EHRs-Data-Harmonization-Platform.git};
    \item Go to the \texttt{App} folder and open the \texttt{App\_v2\_1\_0.R} file with RStudio;
    \item On RStudio, click on Run App;
    \item The Shiny app interface of the EHRs Data Harmonization Platform should appear on RStudio.
\end{enumerate}

If you cannot install RStudio on your computer, you can go to the \texttt{App} folder and run the following command in a Linux shell console: \texttt{Rscript App\_v2\_1\_0.R}

The platform should automatically install the possible missing software libraries and launch the Shiny app on a local IP address.

For example, on a Dell Latitude~3540 personal computer running a Linux Ubuntu 20.04.6 LTS operating system, the previous command generated an output log containing the following information:
\\

\texttt{Listening on http://127.0.0.1:4897}
\\

By opening this http://127.0.0.1:4897 IP address with a traditional web browser such as Mozilla Firefox, it is possible to see and use the Shiny app.

\subsection{An overview of the platform and its features} \label{appsoverview}
\begin{figure}[!httb]
\begin{center}
\includegraphics[width=0.7\textwidth]{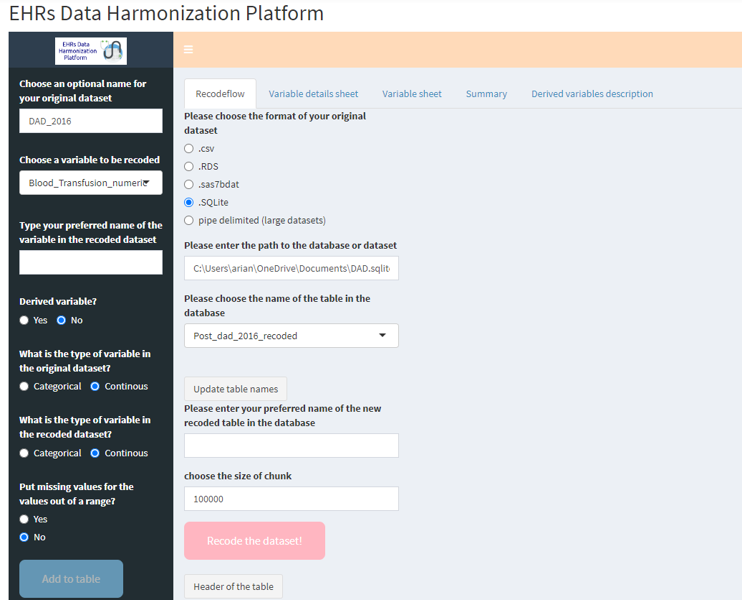}
\caption{An example of a connected SQLite database that enables users to import, curate, and export data in chunks for the efficiency}.
\label{fig:Overview}
\end{center}
\end{figure}

\begin{figure}[!httb]
\begin{center}
\includegraphics[width=0.7\textwidth]{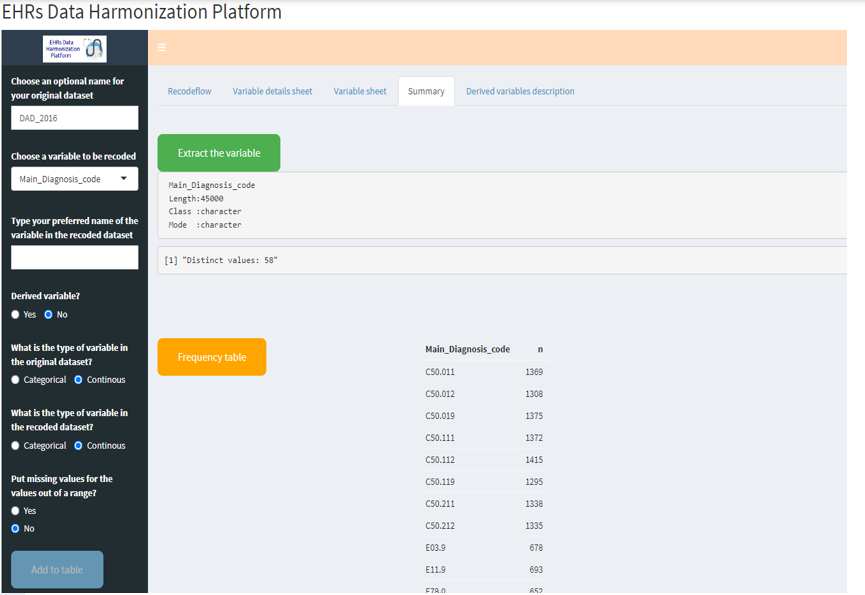}
\caption{The summary tab allows user to extract information about a variable they want to recode.}.
\label{fig:summarytab}
\end{center}
\end{figure}

The shinyapp has 6 main tabs. The “Recodeflow” tab (\autoref{fig:Overview}) is where we connect/upload the non\_curated database and once we used the sidebar panel and updated “variable details sheet” and “variable sheet”, we can determine the output of the recoded dataset and start the recoding process by clicking on the “recoded” data.  One important step in curation is to have information of how a variable really looks in a non-curated database. The “summary” tab (\autoref{fig:summarytab}) allows users to extract information about a variable in the database, see the distribution of different categories and have a better understanding of the variable they wish to recode. Finally, there’s the “derived variables documentation” tab that stores the information of derived variables which use a pre-programmed, custom function such as: the R code of the function and the name of the function. We will illustrate how this tab works in our example in \autoref{sec:reproduciblity} 


\subsection{Renaming a variable and recode the categories}\label{GenderExample}
\begin{figure}[!httb]
\begin{center}
\includegraphics[width=0.7\textwidth]{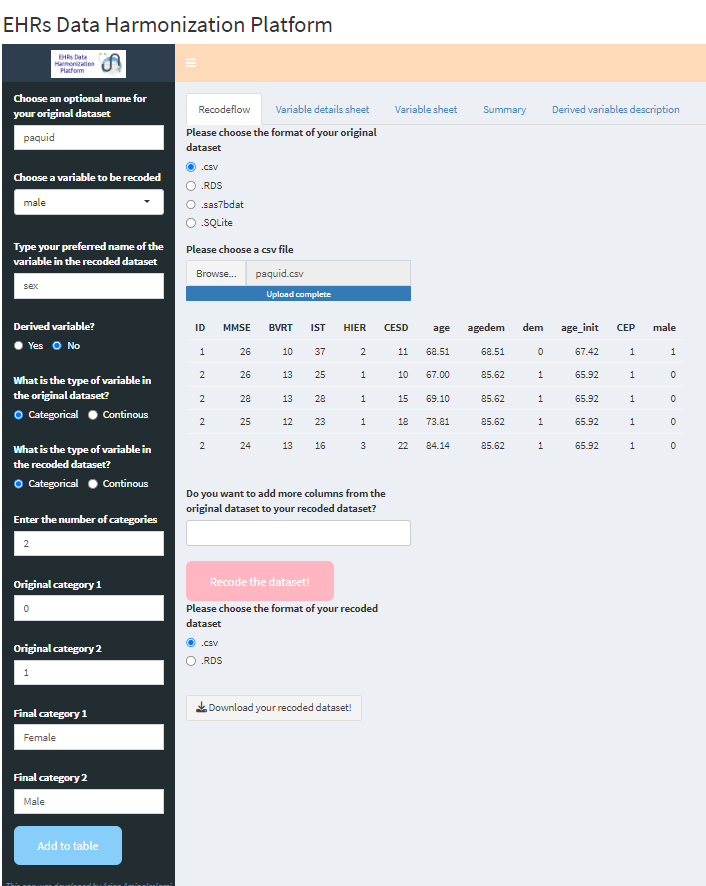}
\caption{An example to rename and recode the categories of a variable}
\label{fig:Gender}
\end{center}
\end{figure}

One of the most common curations in databases is to rename a variable. In our example, there’s a “male” variable in Paquid dataset which gets binary values of 0,1. We want to first rename the variable to “sex” and then recode it so that 0 represents “Female” and 1 would be “Male”. To do so, we should first follow the following initial steps:

\begin{enumerate}
    \item We upload the Paquid dataset by selecting .csv, clicking on “Browse”, and selecting the paquid.csv file
on our computer. This CSV file is available within our GitHub repository (“Data availability” section);
    \item (Optional) We call this dataset Paquid by writing it in the “Choose an optional name for your original dataset” field;
\end{enumerate}
After these preliminary steps, we need to follow these steps: 
\begin{enumerate}
  \setcounter{enumi}{2}
     \item Choose the “male” variable;
     \item type our preferred new name for the variable which is “sex”; 
     \item Choose the original and recoded data type which is Categorical to Categorical;
     \item Enter the number of categories which is 2, and enter how categories should be recoded;
\end{enumerate}
Once all these steps are done, we only need to add the information to the details sheet by clicking on the “add to table” button (\autoref{fig:Gender})

\subsection{Derived variable from one variable, using built-in transformations:  MMSE as category }
\label{sec:MMSE-CATEGORY}

Let us explain a use case where a user would like to map the mini mental state examination~(MMSE) numerical feature into an MMSE category in the Paquid dataset.
In the original Paquid dataset, in fact, the MMSE comes as a number between 0 and 30, with higher scores indicating better cognitive function. 
We can categorize the total scores into ranges, such as:
\begin{itemize}
    \item 0-9: severe cognitive impairment;
    \item 10-17: moderate cognitive impairment;
    \item 18-23: mild cognitive impairment (MCI);
    \item 24-30: normal.
\end{itemize}

Now we would like to create a new categorical feature that indicates these four pieces of information from the MMSE numerical variable.

To do so, we can follow these steps:
\begin{enumerate}
    \item We select \emph{MMSE} in the ``Choose a variable to be recoded'' field;
    \item We choose a name for the recoded variable (\emph{MMSE\_category}) and write it in the ``Type your preferred name of the variable in the recoded dataset'' field;
    \item We enter 4 in the ``Enter the number of categories'' field;
    \item We select No in the ``Derived Variable'' to state ours is not a derived feature;
    \item We select \emph{continuous} in the ``what is the type of variable in the original dataset'' to state that the input MMSE variable is numerical;
    \item We select \emph{categorical} in the ``what is the type of variable in the recoded dataset'' to state that our desired output MMSE category variable is categorical;
    \item We indicate the intervals of the values of the input MMSE variable in the ``Lowerbound~1'', ``Upperbound~1'', ..., ``Lowerbound~4'', and ``Upperbound~4'' fields: 0 and 9, 10 and 17, 18 and 23, and 24 and 30;
    \item We indicate the corresponding new four categories of the MMSE category variable in the ``Final category 1``, ..., ``Final category 4'' fields: severe cognitive impairment, moderate cognitive impairment, mild cognitive impairment, and normal, respectively;
    \item We select 0 for the ``Enter the row number to be deleted'' field;
    \item We click on the ``Add to table'' button;
    \item In the ``Do you want to add more columns from the original dataset to your recoded dataset?'', we select the names of all the original variables;
    \item We click on the ``Recode the dataset'' button;
\end{enumerate}

This is a good example to discuss how recodeflow handles missing values.  recodeflow has a standard approach to handle missing values, by recoding missing data categories values into 3 NA values that are commonly used for most studies:

\begin{itemize}
    \item NA(a) = 'not applicable'
    \item NA(b) = 'missing'
    \item Na(c) = 'not asked'
\end{itemize}

\begin{figure}[!httb]
\begin{center}
\includegraphics[width=0.7\textwidth]{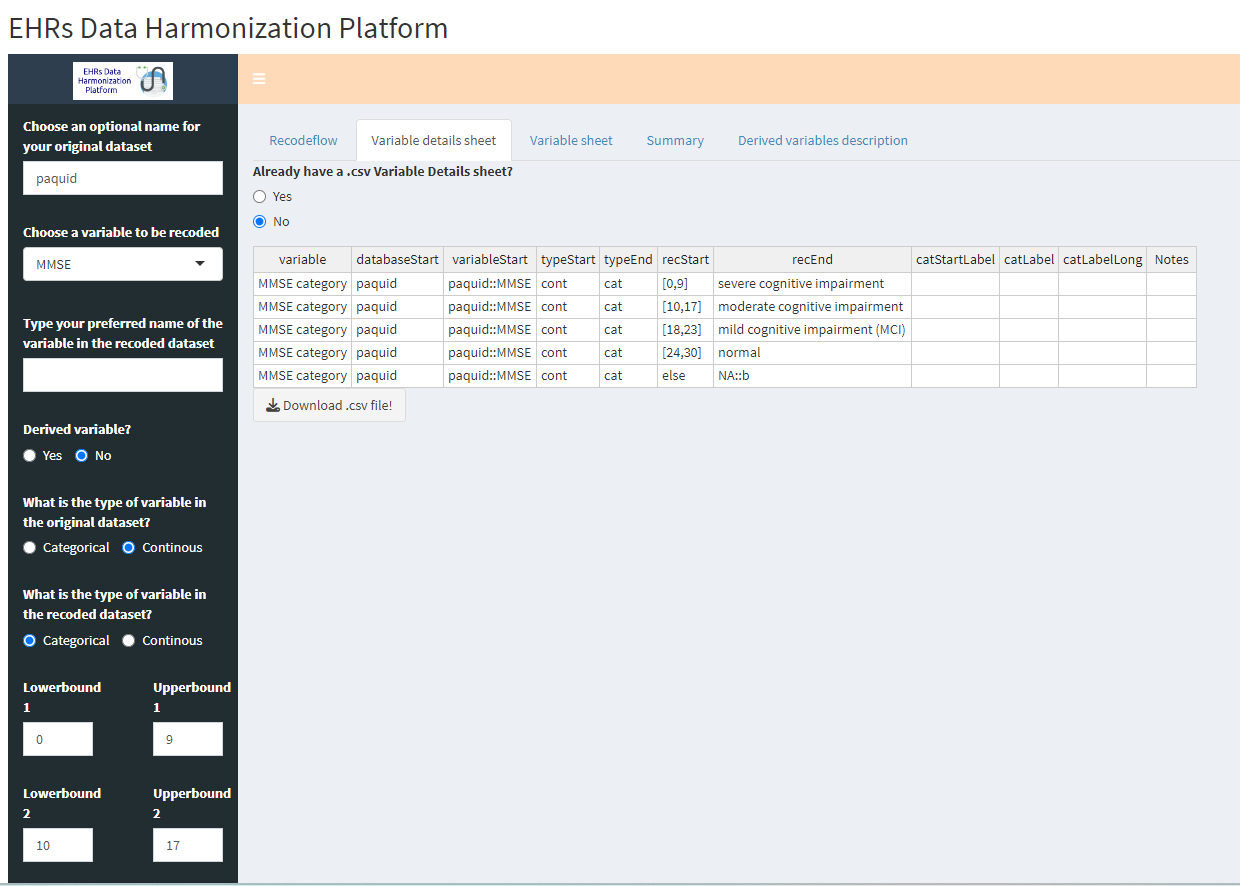}
\caption{An example showing how to categorize a continuous variable and how they appear in the details sheet.}
\label{fig:mmse}
\end{center}
\end{figure}

In this example, we will specify that any categories other than what we specified, should be categorized as NA(b) or missing. 
\autoref{fig:mmse} shows the steps we followed for this example and how the missing values are handled in the details sheet.  Note that labels and notes are not mandatory fields of the details sheet, so there are no inputs for them on the app and users can modify them on the table’s relevant cell if they prefer to. 

\subsection{Derived variable from two variables that needs custom functions: MMSE and CEP }
\label{sec:MMSE-CEP}
In this last use case, we will see an example of a derived variable with a pre-programmed, custom function. \textit{recodeflow} supports use of any custom functions as long as the variable can be calculated on a per row basis.
Let us suppose we would like to have a feature that indicates both the MMSE category and the education level of the patients, which is recorded in the dataset through the CEP factor.
CEP = 1 means that the patient graduated from primary school, and CEP = 0 means she or he did not.

It might be useful to have a derived feature that indicates both the MMSE category and the CEP status~(\autoref{tab:MMSE-CEP-VALUES}).

\begin{table}[!httb] \scriptsize
 \begin{tabularx}{\textwidth}{lllll}
& \textbf{severe cognitive} & \textbf{moderate cognitive} & \textbf{mild cognitive} & \textbf{normal}  \\
& \textbf{impairment} & \textbf{impairment} & \textbf{impairment} & \textbf{condition}  \\
\midrule
\textbf{graduated}     & severe cognitive     & moderate cognitive     & mild cognitive      & normal condition     \\
 &  impairment and graduated     &  impairment and graduated     &  impairment and graduated     &  and graduated     \\
 \midrule
\textbf{non-graduated} & severe cognitive  & moderate cognitive  & mild cognitive impairment  & normal condition \\
&  impairment and non-graduated &  impairment and non-graduated &  non-graduated & and non-graduated \\
\end{tabularx}
\caption{\textbf{Table of the MMSE-CEP values}.
Each patient can have one of these eight values for the derived variable MMSE-CEP
}
\label{tab:MMSE-CEP-VALUES}
\end{table}

To generate this derived feature, we can use the MMSE\_category variable we produced at the previous step, and we need to create a new recoded feature for CEP, too.
As per rules of \texttt{recodeflow}, a derived variable can be created only from recoded variables.

We follow the steps described in the previous section and generate a recoded variable called \emph{CEP\_bin}, which has value \emph{graduated} when CEP equals to 1 and has value \emph{non-graduated} when CEP equals to 0.

\begin{figure}[!httb]
\begin{center}
\includegraphics[width=0.7\textwidth]{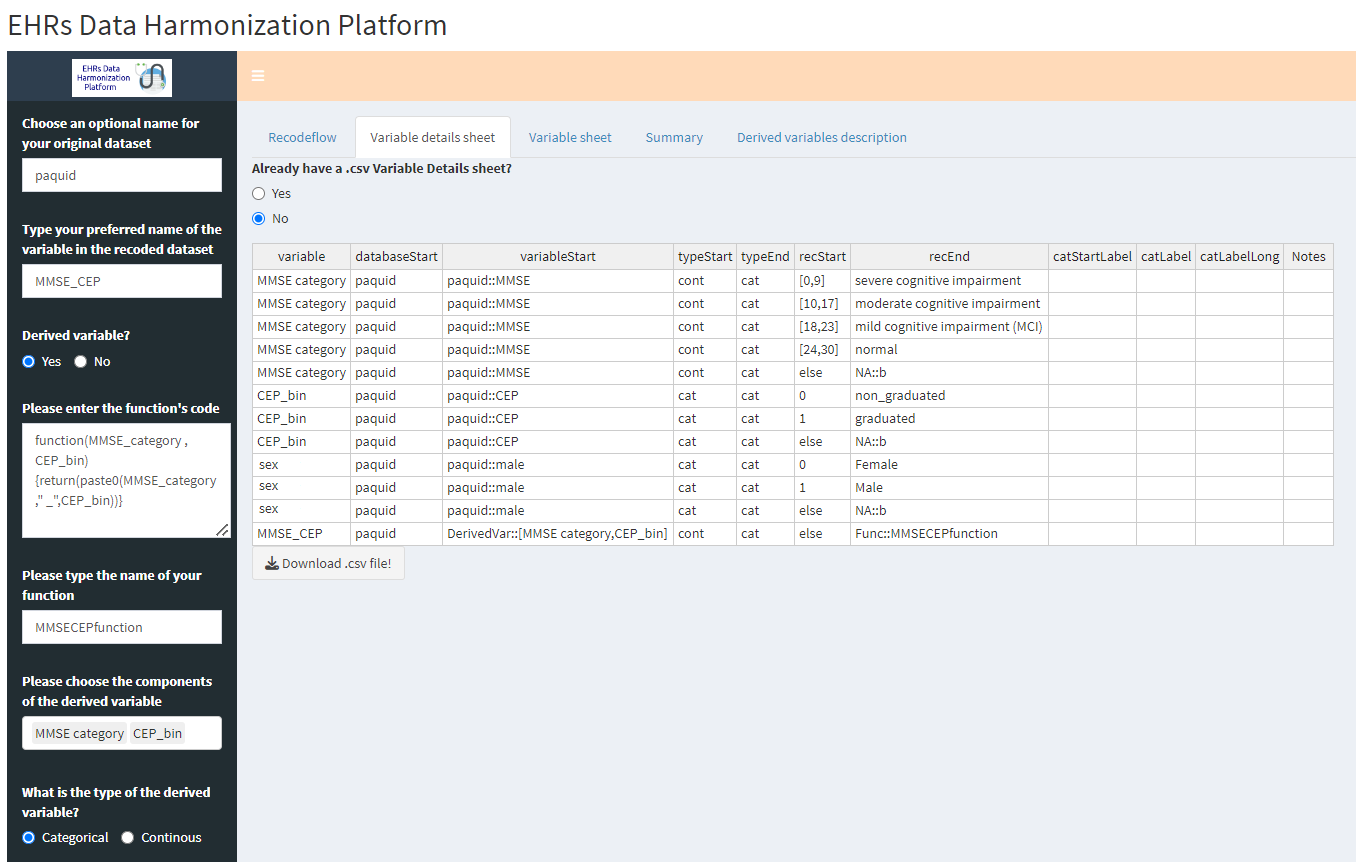}
\caption{Steps needed to create a variable derived from more than a variable and the updated details sheet.}
\label{fig:DERIVED-SCREENSHOT}
\end{center}
\end{figure}

To create a new derived feature based on MMSE\_category and CEP\_bin, we follow these steps:
\begin{enumerate}
    \item We write \emph{MMSE-CEP} in the ``Type your preferred name of the variable in the recoded dataset'' field;
    \item In the ``Derived Variable?'' field, we select Yes;
    \item In the ``Please enter the function's code'' field, we insert R code of the function that needs to create the new feature. In our case: \\
    \texttt{function(MMSE\_category, CEP\_bin)\{return(paste0(MMSE\_category,"\_",CEP\_bin))}\};
    \item We can write \emph{MMSECEPfunction} in the ``Please type the name of your function'' field;
    \item In the ``Please choose the components of the derived variable`` field, we select \emph{MMSE\_cat} and {CEP\_bin};
    \item We select \emph{categorical} in the ``what is the type of the derived variable?'' field;
    \item We select 0 in the ``enter the row number to be deleted'' field;
    \item We click on ``Add to table'';
    \item In the central lower menu, we select all the features of the dataset in the ``Do you want to add more columns from the original dataset to your recoded dataset?'' field;
    \item We finally click on ``Recode this dataset!'';
    \item At this point, the new column \emph{MMSE-CEP} should appear in the header of the recoded dataset;
    \item We click on ``Download the recoded dataset!'' and save the dataset file in CSV format.
\end{enumerate}
\autoref{fig:DERIVED-SCREENSHOT} shows the steps as well as the updated details sheet.

\begin{figure}[!httb]
\begin{center}
\includegraphics[width=0.89\textwidth]{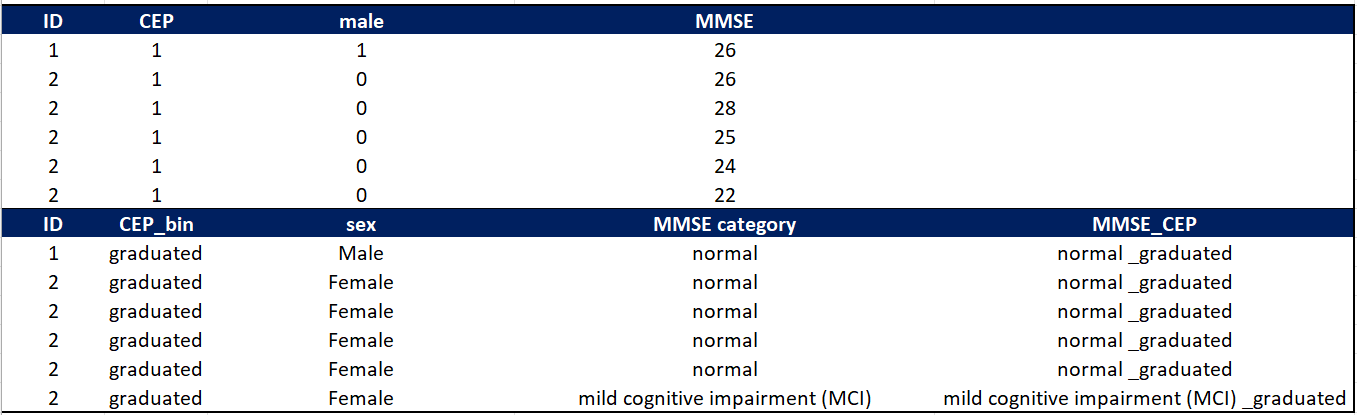}
\caption{The comparison between the original (top) and downloaded curated version of Paquid dataset (bottom).}
\label{fig:comparison}
\end{center}
\end{figure}

At the end of this procedure, we obtain a new version of the dataset containing the MMSE-CEP variable that puts into relationship the mental health category and the education level of each patient. You can see the comparison between the original and curated data set in \autoref{fig:comparison}.

\subsection{Reproducibility of the work}
\label{sec:reproduciblity}

\begin{figure}[!httb]
\begin{center}
\includegraphics[width=0.7\textwidth]{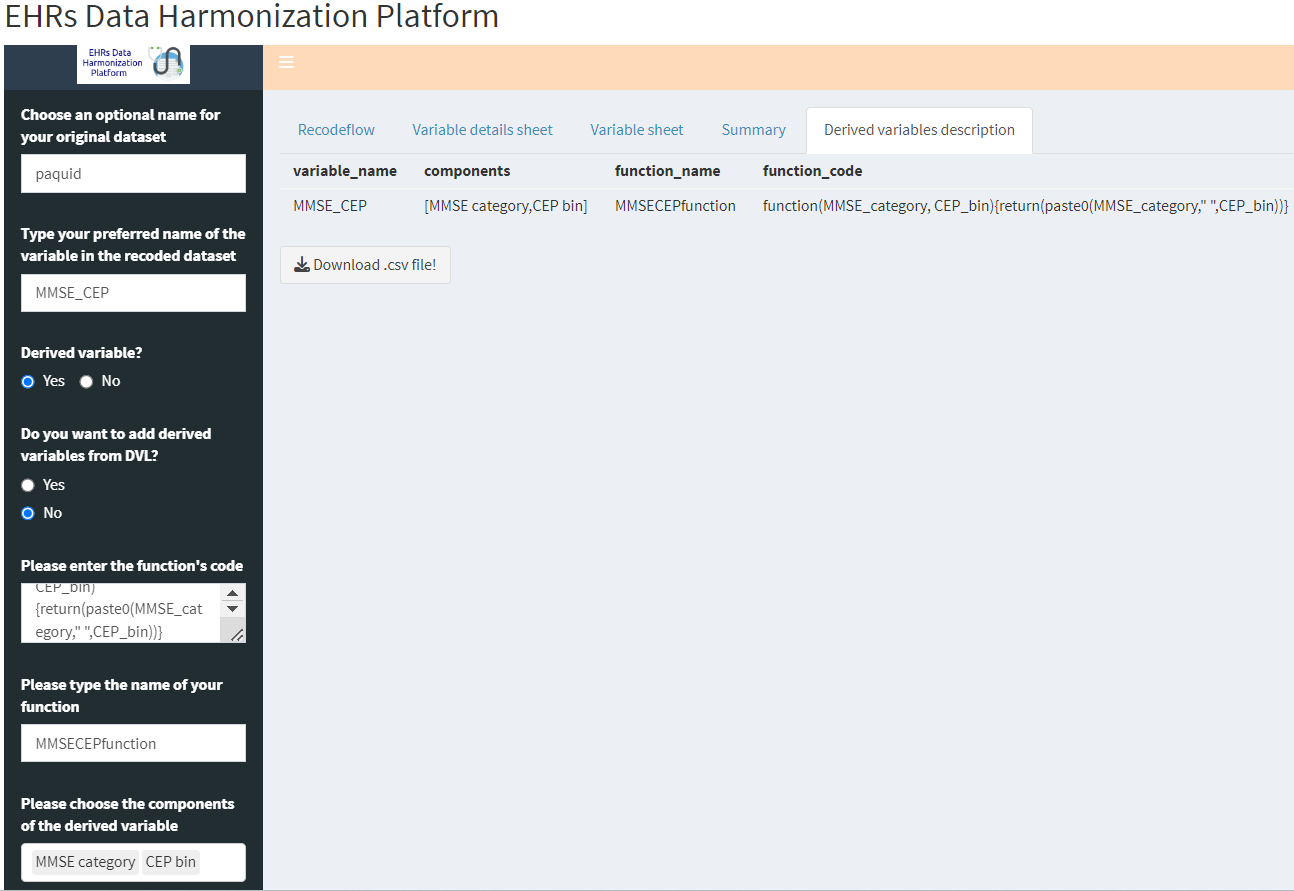}
\caption{As soon as a derived variable (that needs a custom function) is added to the details sheet, the variables’ name, components, function name and function code will be saved in “Derived variable description” tab. This spread sheet can be downloaded and used to create/update a derived variables library.}
\label{fig:derived documentation}
\end{center}
\end{figure}

The main contribution of our EHRs data curation platform is the reproducibility and shareability of the result. In the “derived variables description” tab, valuable information about the derived variable we created in \autoref{sec:MMSE-CEP} is stored (\autoref{fig:derived documentation}). Also, the details sheet that we created has essentials information about the changes we made to the dataset. Both of these spreadsheets are available to download and share with others. Other users only need to upload the details sheet, connect to a “derived variables library” which is a repository of all derived variables that are created and verified before by other researchers (with the same structure of the spreadsheet available in the “derived variables description” tab), select their required derived variables and reproduce a curated dataset. Also, users can modify and update an existing work with this approach.
To reproduce a work, after connecting/uploading your original database: 

\begin{enumerate}
    \item In the details sheet tab, an existing details sheet will be uploaded.
    \item In the sidebar panel, after selecting “yes” for derived variable, we choose to select a derived variable from a derived variable library (DVL), we enter the path to it, and select our preferred derived variables.
    \item We click on the “recode the dataset” button and we can reproduce a curation work.
\end{enumerate}

\section{Discussion and conclusions}
\label{sec:DISCUSSION-AND-CONCLUSIONS}

Electronic health records contain data of patients which, even if not collected for scientific purposes, can pave the way to important clinical discoveries.
These data, in fact, can be analyzed through computational programs using machine learning and biostatitics, which can reveal data trends and outcomes that otherwise would not be noticeable by medical doctors.
Data of EHRs and surveys, however, seldom come in a format that is ready for computational analyses: most of the times, these datasets need the preprocessing phases of variable harmonization and feature derivation before serving as input to a computational intelligence pipeline.

These preprocessing steps are usually done manually or with \emph{ad hoc} scripts by researchers and users worldwide, who can accidentally make mistakes that could invalidate the whole scientific project.
To solve this problem, we present here our EHRs Data Harmonization Platform: an easy-to-use Shiny app and R library which can be utilized for free by anyone in the world for data harmonization and derivation.

In addition to performing these preprocessing steps automatically, our platform also provides an user-friendly user interface, that can exploited easily by any user, even those without a computational background.
Moreover, our platform also automatically produces the documentation regarding these data transformations, through a document called \emph{variable detail sheet}.

Regarding limitations, we need to report that EHRs Data Harmonization Platform has some difficulties handling large datasets in CSV, RDS or SAS formats; we recommend users having large datasets to convert them in SQLite, which is the only format that currently can handle them efficiently.
We broadly consider \emph{large} all the datasets greater or equal to 20 gigabytes~(GB).

To the best of our knowledge, no other software library existing in the scientific literature provides similar assets.
In the future, we plan to develop a Python version of our platform.

\newpage
\clearpage

\section*{Additional information}
\paragraph{List of abbreviations}
BVRT:~Benton Visual Retention Test.
CCHS:~Canadian Community Health Service.
CESD:~Center for Epidemiological Studies Depression Scale.
CIHR:~Canadian Institutes of Health Research.
CRAN:~Comprehensive R Archive Network.
CSV:~comma-separated values.
DAD:~Discharge Abstract Database.
dem:~dementia.
DVL:~derived variable library.
EHR:~electronic health record.
ID:~identifier.
IP:~internet protocol.
IST:~Isaacs Set Test.
MMSE:~Mini-Mental State Examination.
NACRS:~National Ambulatory Care Reporting System.
ODB:~Ontario Drug Benefit.
OHDP:~Ontario Health Data Platform.
OHIP:~Ontario Health Insurance Plan.
SAS:~Statistical Analysis System.
URL:~Uniform Resource Locator.

\paragraph{Conflict of interest}
The authors declare they have no conflict of interest.

\paragraph{Funding}
This study is part of the Broad and Deep Longitudinal Analysis in Neurodegenerative Disease~(BRAIN) project and is supported by the Canadian Institutes of Health Research (CIHR), in partnership with CIHR's Institute of Aging and CIHR's Institute of Neuroscience, Mental Health and Addiction~(CIHR Funding Reference Number: BDO 148341).
This study also was funded by the European Union – Next Generation EU programme, in the context of The National Recovery and Resilience Plan, Investment Partenariato Esteso PE8 ``Conseguenze e sfide dell'invecchiamento'', Project Age-It (Ageing Well in an Ageing Society) and also partially supported by Ministero dell'Universit\`a e della Ricerca of Italy under the ``Dipartimenti di Eccellenza 2023-2027'' ReGAInS grant assigned to Dipartimento di Informatica Sistemistica e Comunicazione at Università di Milano-Bicocca.
The funders had no role in study design, data collection and analysis, decision to publish, or preparation of the manuscript.

\paragraph{Ethics approval and consent to participate}
Ethics approval and consent to participate from the patients to the Paquid study were collected by the dataset original curators~\cite{letenneur1994incidence}.

\paragraph{Acknowledgments}
The authors thank Dorsa Ghahramani~(University of Toronto) and Douglas Manuel~(the Ottawa Hospital) for their help.

\paragraph{Data availability}
The Paquid dataset used in this study is publicly available within the \texttt{lcmm} R software package and on our GitHub repository at: \\
\url{https://github.com/ArianAminoleslami/EHRs-Data-Harmonization-Platform/blob/main/data/paquid.csv}

More information about the Paquid dataset can be found in the study by Luc Letenneur and colleagues~\cite{letenneur1994incidence} and on CRAN at: \url{https://search.r-project.org/CRAN/refmans/lcmm/html/paquid.html}

\paragraph{Software availability}
Our R package source code is publicly available under the GPL-3.0 license on GitHub at: \\
\url{https://github.com/ArianAminoleslami/EHRs-Data-Harmonization-Platform}

Our Shiny app is also available via web browser under the GPL-3.0 license and can be accessed through internet browser at: \\
\url{https://poxotn-arian-aminoleslami.shinyapps.io/Arian/}

This manuscript refers to the release~v1.0.1 of our platform.

\footnotesize
\bibliographystyle{unsrt}
\bibliography{bibliography_recodeflow_file.bib}

\normalsize
\newpage
\clearpage
\beginsupplement
 
\section*{Supplementary information}
\label{sec:SUPPLEMENTARY-INFORMATION}

\begin{figure}[!httb]
\begin{center}
\includegraphics[width=0.9\textwidth]{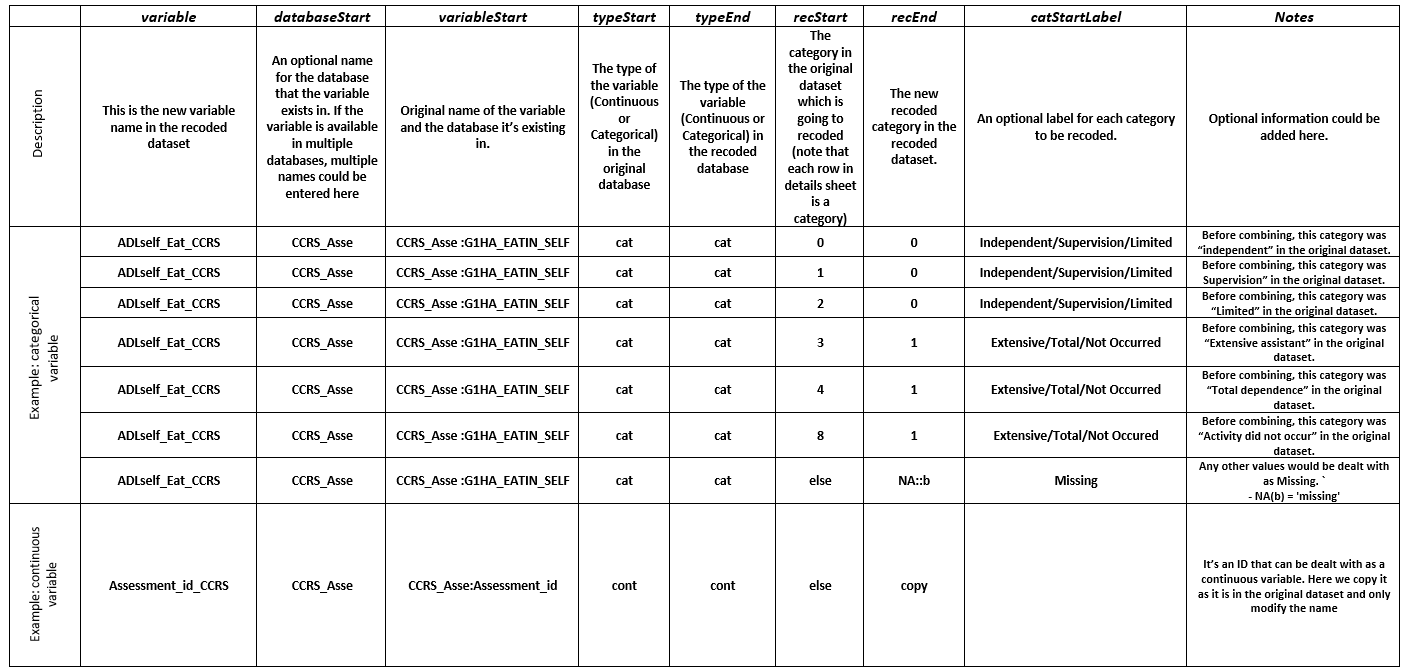}
\caption{The description of information stored in a “variable details sheet” with some examples. Each row of the variable details sheet represents a category which is going to be recoded.
\label{fig:DESCRIPTION}}
\end{center}
\end{figure}

\begin{figure}[!httb]
\begin{center}
\includegraphics[width=0.9\textwidth]{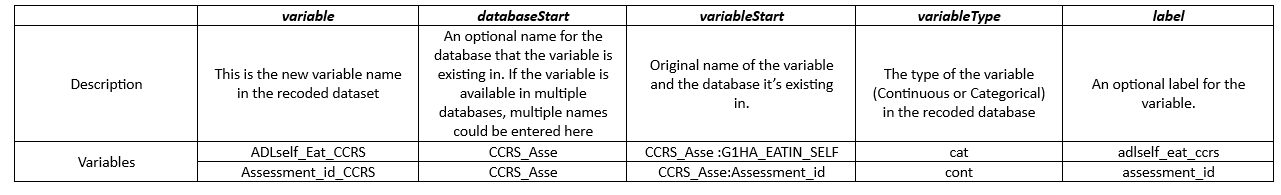}
\caption{An example of a ``variable sheet'', and the description of its columns. The number of rows in this spreadsheet represents the total number of variables picked for curation.
\label{fig:VARIABLE-SHEET}}
\end{center}
\end{figure}

\end{document}